\newcommand{\Hil}{{\mathcal H}}
\newcommand{\Ell}{{\mathcal L}}
\newcommand{\Prho}{{\mathcal P}}
\newcommand{\Boltz}{ k_{\rm\scriptscriptstyle B}}

\newcommand{\Tr}{{\rm Tr}}

\newcommand{\Ker}{{\rm Ker}}
\newcommand{\D}{{\bf \hat D}}

\newcommand{\ddt}[1]{{\frac{\displaystyle {\rm d}#1}{\displaystyle {\rm d}t}}}

\newcommand{\cov}[2]{{\langle\Delta #1\Delta #2\rangle}}
\newcommand{\Otimes}{{ \otimes }}

\documentclass[prl,aps,twocolumn,showpacs,preprintnumbers,amsmath,amssymb]{revtex4}

\begin{document}

\title{ Nonlinear extensions of Schr\"odinger--von\,\,Neumann quantum dynamics: \\ a list of conditions for compatibility with thermodynamics}
\author{ Gian Paolo Beretta }
\email{beretta@unibs.it}
\affiliation{Istituto Nazionale di Fisica della Materia,
Universit\`a di Brescia, via Branze 38, 25123 Brescia, Italy}
\date{\today}
\begin{abstract} We propose a list of conditions that consistency with thermodynamics imposes on linear and nonlinear generalizations of standard unitary quantum mechanics that assume a set of true quantum states without the restriction $\rho^2=\rho$ even for strictly isolated systems and that are to be considered in experimental tests of the existence of intrinsic (spontaneous) decoherence at the microscopic level. \end{abstract}

\maketitle

Understanding and predicting 'decoherence' is important in future applications involving nanometric devices, fast switching times, clock synchronization,  superdense coding, quantum computation, teleportation, quantum cryptography, etc. where entanglement structure and dynamics play a key role \cite{decoherence}. In the last three decades it has also been central in exploring possible limitations to the validity of standard unitary quantum mechanics (QM), by studying a variety of linear and nonlinear extensions that have been advocated by several authors on a variety of conceptual grounds \cite{extensions}. It has been suggested \cite{Domokos} that long-baseline neutrino oscillation experiments  may provide means of testing the existence of spontaneous decoherence at the microscopic level and the validity of linear and nonlinear extensions of the Schr\"odinger--von\,\,Neumann equation of motion of QM, thus prompting a renewed interest on such extensions \cite{Domokos,Czachor,Gheorghiu}.

The aim of this Letter is to list the main conditions that must be imposed and checked on linear and nonlinear extensions of QM which assume an augmented set of true quantum states described by state operators $\rho$ without the restriction $\rho^2=\rho$. The reasoning and framework proposed here should provide useful guidance also to current efforts to define general measures of entanglement \cite{Yukalov}.

The conditions proposed here form a very restrictive set. Yet, at least one possible extension has been proved to satisfy them all\cite{Beretta}, with mathematics that has been partially rediscovered recently by researchers in different contexts and fields \cite{Gheorghiu,Englman,Aerts}.

\smallskip\noindent{\bf 1.\ Causality. Forward and backward in time.\\}\label{causality} 
We consider the set $\Prho$ of all linear, hermitian, nonnegative-definite, unit-trace  operators $\rho$ (without the restriction $\rho^2=\rho$) on the standard QM Hilbert space $\Hil$ associated with a strictly isolated system\cite{isolated}. Every solution of the equation of motion, i.e., every trajectory $u(t,\rho)$ passing  at time $t=0$ through state $\rho$ in $\Prho$, should lie entirely in $\Prho$ for all times $t$, $-\infty<t<+\infty$. This strong causality condition is nontrivial and demanding both from the conceptual and the technical mathematical points of view.

\smallskip\noindent{\bf 2.\ Conservation of energy and other invariants.}\label{energy}
The value of the energy functional $e(\rho)=\Tr(\rho H)$, where $H$ is the standard QM Hamiltonian operator associated with the isolated system [$H\ne H(t)$],
must remain invariant along every trajectory. If $\Hil$ is the Fock space of an isolated system consisting of $M$ types of elementary constituents (e.g., atoms and molecules if chemical and nuclear reactions are inhibited; or atomic nuclei and electrons for modelling chemical reactions) each with a number operator $N_i$ ($[H,N_i]=0$ and $[N_i,N_j]=0$), then also the value of each number-of-constituents functional $n_i(\rho)=\Tr(\rho N_i)$ must remain invariant along every trajectory.  Depending on the type of system, there may be other time-invariant functionals, e.g., the total momentum components $p_j (\rho)=\Tr(\rho P_j)$, with $j=x,y,z$, for a free particle (in which case Galileian invariance must also be verified, for $[H,P_j]=0$ and $[P_i,P_j]=0$). In what follows, we denote by $g_i(\rho)=\Tr(\rho G_i)$ the set of non-Hamiltonian time-invariant functionals, if any, with $[H,G_i]=0$ and $[G_i,G_j]=0$ (clearly, $H$ and the $G_i$'s have a common eigenbasis that we denote by $\{|\psi_\ell\rangle\}$).

\smallskip\noindent{\bf 3.\ Standard QM unitary evolution of $\rho^2=\rho$ states.\\}\label{standardQM}
Unitary time evolution of the states of QM according to the Schr\"odinger
equation of motion must be compatible with the more general dynamical law.  These trajectories, passing through any state
$\rho$ such that $\rho^2=\rho$ and entirely contained in the state domain of
QM, must be solutions also of the extended dynamical law. Because the states of
QM are extreme points of the state domain $\Prho$, the
trajectories of QM must be boundary solutions (limit cycles) of the extended dynamical law.

In general, any extended dynamical equation may be written in the form
\begin{eqnarray}\label{eqofmotion}
&&{\displaystyle \ddt{\rho} = - \frac{i}{\hbar}}[H,\rho] +D_M \\&&{\rm with}\ D_M=\D_M(\rho,H,G_i,\dots) \ ,
\end{eqnarray}
where operator $D_M$ represents the {\it dissipative} part of the equation of motion and may depend linearly and/or nonlinearly (through superoperator $\D_M$) on the state operator $\rho$, on the Hamiltonian $H$, on the linear operators $G_i$ associated with the other time invariants (if any), as well as on the structure and the number $M$ of elementary constituents of the system. 
Like the Schr\"odinger--von\,\,Neumann term, also the dissipative term should not be responsible for rates of change of any of the invariant functionals $\Tr(\rho)$, $e(\rho) $, $g_i(\rho) $ and, therefore, 
\begin{equation}
\Tr D_M=0\qquad\Tr D_M H=0\qquad\Tr D_M G_i=0\ .
\end{equation}

If the complete dynamics preserves the feature of uniqueness of
solutions throughout the state domain $\Prho$, then pure states
can only evolve according to the Schr\"odinger equation of motion and, therefore, $\D_M(\rho,H,G_i,\dots)=0$ when $\rho^2=\rho$. This feature that may be responsible for hiding the presence of deviations from QM in experiments where the isolated system is prepared in a pure state.
It also implies that no trajectory can enter or leave the state domain of
QM. Thus, by continuity, there must be trajectories that approach
indefinitely these boundary solutions (of course, this can only
happen backward in time, as $t\to -\infty$, for otherwise the entropy of the isolated system would decrease in forward time). 

\smallskip\noindent{\bf 4.\ Conservation of effective Hilbert space dimensionality.} 

Unitary dynamics [Eq.\ (\ref{eqofmotion}) with $D_M=0$] would maintain unchanged all the eigenvalues of $\rho$ and therefore cannot satisfy Condition 5 below \cite{unitary}. Instead, we only require that the dynamical law maintains zero the initially zero eigenvalues of $\rho$ and, therefore, conserves the cardinality of the set of zero eigenvalues, $ \dim\Ker(\rho)$. In other words, if the isolated system is prepared in a state that does not 'occupy' the eigenvector $ |\psi_\ell\rangle $ of $H$ (and the $G_i$'s), i.e., if $\rho(0)|\psi_\ell\rangle =0$ (so that $|\psi_\ell\rangle $ is also an eigenvector of $\rho$ corresponding to a zero eigenvalue), then such energy eigenvector remains 'unoccupied' at all times, i.e., $\rho(t)|\psi_\ell\rangle =0$. 

This condition preserves an important feature that allows remarkable model simplifications within QM: the dynamics is fully equivalent to that of a model system with Hilbert space $\Hil'$ (a subspace of $\Hil$) defined by the linear span of all the $|\psi_\ell\rangle$'s such that $\rho(t)|\psi_\ell\rangle \ne 0$ at some time $t$ (and, hence, by our condition, at all times). The relevant operators $X'$ on $\Hil'$ ($\rho'$, $H'$, $G'_i$, \dots) are defined from the original $X$ on $\Hil$ ($\rho$, $H$, $G_i$, \dots) so that $\langle \alpha_k|X'|\alpha_\ell\rangle =\langle \alpha_k|X|\alpha_\ell\rangle$ with $|\alpha_k\rangle$ any basis of $\Hil'$.

It is also consistent with recent experimental tests \cite{exp1} that rule out, for pure states, deviations from linear and unitary dynamics and confirm that initially unoccupied eigenstates cannot spontaneously become occupied. This fact adds nontrivial experimental and conceptual difficulty to the problem of designing a fundamental test of QM, capable of ascertaining whether decoherence originates from uncontrolled interactions with the environment due to the practical impossibility of obtaining strict isolation, or else it is a more fundamental intrinsic feature of microscopic dynamics requiring an extension of QM.  In the latter case, this condition will preserve within the extended theory the exact validity of all the remarkable successes of QM.

\smallskip\noindent{\bf 5.\ Entropy nondecrease. Irreversibility.}\label{irreversibility}
The principle of nondecrease of entropy \cite{nonextensiveS} for an isolated system must be satisfied, i.e., the rate of
change of the entropy functional $-\Boltz\Tr(\rho\ln\rho)$ must be nonnegative along every trajectory, $-\Boltz\Tr [u(t,\rho)\ln u(t,\rho)]\ge
-\Boltz\Tr(\rho\ln\rho)$.

\smallskip\noindent{\bf 6.\ Stability and uniqueness of the thermodynamic equilibrium states. Second law.}\label{secondLaw}
A state operator $\rho$ of the isolated system represents an {\it equilibrium state} if ${\rm
d}\rho/{\rm d}t=0$.
For each given set $(\tilde e, {\bf \tilde g})$ of feasible values of the energy functional
$e(\rho)$ and the other time-invariant functionals $g_i(\rho)$, if any,
among all the equilibrium states that the dynamical law may admit there
must be one and only one which is {\it globally stable}\cite{stability}.  

This stable equilibrium state
must be that of equilibrium thermodynamics
and, therefore, of the form 
\begin{equation}\label{stableeq}
\rho_{\rm e}  =\frac{ \exp[-\beta(\tilde e, {\bf \tilde g}) H+\sum_i\nu_i(\tilde e, {\bf \tilde g}) G_i]}{\Tr \exp[-\beta(\tilde e, {\bf \tilde g}) H+\sum_i\nu_i(\tilde e, {\bf \tilde g}) G_i] }\ ,
\end{equation}
where $G_i$ are defined above. Of course, 
states given by Eq.\ (\ref{stableeq}) are solutions of the constrained maximization problem
\begin{subequations}\label{maxproblem}\begin{eqnarray}
&& {\rm max}\ -\Boltz\Tr(\rho\ln\rho)\ {\rm subject\ to\ }\\
&& \Tr(\rho)=1,\ \Tr(\rho H)=\tilde e,\ \Tr(\rho G_i)=\tilde g_i,\ {\rm and\ } \rho \ge 0 \ .
\end{eqnarray}\end{subequations}
and reduce to the canonical equilibrium states 
$\rho_{\rm e}  = \exp(-\beta H)/\Tr \exp(-\beta H)$
when $G_i=g_iI$ for all $i$'s (with $g_i$ scalars and $I$ the identity on $\Hil$), and to the microcanonical state $\rho_{\rm e}=I/\dim\Hil$ if also $H=eI$ (and $\dim\Hil <\infty$).

As discussed in Ref.\ \cite{JMathPhys}, the entropy functional is not a Lyapunov function, even
if, in a strict sense that depends on the continuity and the
conditional stability of states $\rho_{\rm e}$,
it does provide a criterion for the local stability of these states.
In addition to this, the second law requires however that no 
other equilibrium state of the dynamical law be
be globally stable\cite{stability,unitary}.

Consider the noteworthy family of states 
\begin{equation}
\rho_{\rm nd}  =\frac{ B\exp[-\beta(\tilde e, {\bf \tilde g})  H  +\sum_i\nu_i(\tilde e, {\bf \tilde g})  G_i ] B}{\Tr B\exp[-\beta(\tilde e, {\bf \tilde g})  H  +\sum_i\nu_i(\tilde e, {\bf \tilde g})  G_i ] }\ ,
\end{equation}
where $ B$ is any given idempotent operator $ B^2= B$. This family, which
includes pure states [$\Tr(B)=1$], maximizes the entropy [Prob.\ (\ref{maxproblem})] subject to the additional constraint $ \rho=B\rho B$ for the given $B$. All eigenvalues of $\rho_{\rm nd}$ must remain invariant (otherwise the entropy would decrease) and the state is equilibrium if $[B,H]=0$ or otherwise it evolves unitarily (limit cycle) with $B(t)=\exp(-iHt/\hbar)B(0) \exp(iHt/\hbar)$. They have a thermal-like distribution (positive and negative temperatures) over a finite number [$\Tr(B)$] of 'occupied' eigenvectors. Because entropy cannot decrease and $-\Boltz\Tr(\rho\ln\rho)$ is an $S$-function, they are conditionally locally stable equilibrium states or limit cycles \cite{JMathPhys}. For them not to be globally stable, as required by the second law, it suffices that the extended dynamics imply that at least one state  of equal energy and other invariants (not necessarily neighboring nor with the same kernel) evolves towards higher entropy than $\rho_{\rm nd}$.

\smallskip\noindent{\bf 7.\ Non-interacting subsystems. Separate energy conservation.}\label{separateEnergyConservation}
For an isolated system composed of two distinguishable subsystems $A$ and $B$ with associated
Hilbert spaces $\Hil^A$ and $\Hil^B$, so that the Hilbert space of the system is
$\Hil=\Hil^A\Otimes\Hil^B$, if the two subsystems are non-interacting, i.e., the
Hamiltonian operator $H = H_A\Otimes I_B + I_A\Otimes H_B$, then the functionals
$\Tr[(H_A\Otimes I_B)\rho]= \Tr_A(H_A\rho_A) $ and $\Tr[(I_A\Otimes H_B)\rho]= \Tr_B(H_B\rho_B)$ represent the
energies of the two subsystems and must remain invariant along every trajectory, even if the states of $A$ and $B$ are correlated, i.e., even if $\rho\ne \rho_A\Otimes\rho_B$. Of course, $\rho_A=\Tr_B(\rho)$, $\rho_B=\Tr_A(\rho)$, $\Tr_B$ denotes the partial trace over $\Hil^B$ and $\Tr_A$
the partial trace over $\Hil^A$.

\smallskip\noindent{\bf 8.\ Independent states. Weak separability. Separate entropy nondecrease.}\label{separateEntropyConservation}
Two distinguishable subsystems $A$ and $B$ are in independent
states if the state operator $\rho=\rho_A\Otimes\rho_B$. 
For any given $\rho$, let us define the idempotent operator $B$ obtained from $\rho$ by substituting in its spectral expansion each nonzero eigenvalue with unity \cite{operatorB} and
the {\it entropy operator} $S =-\Boltz B\ln\rho$ (always well-defined). For independent states, $S= S_A\Otimes I_B +
I_A\Otimes S_B= -\Boltz [B_A\ln\rho_A\Otimes I_B + I_A\Otimes
B_B\ln\rho_B ]$. For permanently non-interacting subsystems, every
trajectory passing through a state in which the subsystems are in
independent states must proceed through independent
states along the entire trajectory, i.e., when two uncorrelated systems
do not interact with each other, each must evolve in time
independently of the other.

In addition, if at some instant of time two subsystems $A$ and
$B$,  not necessarily non-interacting, are in independent states,
then the instantaneous rates of change of the subsystem's entropy
functionals $-\Boltz\Tr (\rho_A\ln\rho_A)$ and
$-\Boltz\Tr (\rho_B\ln\rho_B)$ must both be nondecreasing in time.

\smallskip\noindent{\bf 9.\ Correlations, entanglement and locality. Strong separability.}\label{strongSeparability}
Two non-interacting subsystems $A$ and $B$ initially in correlated and/or entangled
states  (possibly due to a previous interaction that has then been
turned off) should in general proceed in time towards less correlated and entangled
states.  In any case, in order for the dynamics
not to generate locality problems, i.e., faster-than-light
communication between noninteracting subsystems (even if in
entangled or correlated states), entanglement and correlations must not increase in the absence of interactions. In other words, when subsystem
$A$ is not interacting with subsystem $B$, it should never be
possible to influence the local observables of $A$ by acting only
on the interactions within $B$, such as switching on and off
parameters or measurement devices within $B$. 

This however does not mean
that existing entanglement and/or correlations between $A$ and $B$ established by past
interactions should have no influence whatsoever on the time
evolution of the local observables of either $A$ or $B$. In
particular, there is no physical reason to request that two different
states $\rho$ and $\rho'$ such that $\rho'_A=\rho_A$ should evolve
with identical local dynamics (${\rm d}\rho'_A /{\rm d}t ={\rm d}\rho_A /{\rm
d}t$) whenever $A$ does not interact with $B$, even if entanglement and/or correlations in state $\rho$ differ from those in state $\rho'$. Rather, the two local evolutions should be different until spontaneous decoherence (if any) will have fully erased memory of the entanglement and the correlations established by the past interactions now turned off. In fact, this may be a possible experimental scheme to detect spontaneous decoherence.

Compatibility with the predictions of QM about the generation of quantum entanglement between
interacting subsystems that emerge through the
Schr\"odinger-von\,\,Neumann term $-i[H,\rho]/\hbar$ of Eq.\ (\ref{eqofmotion}), requires that the dissipative term $D_M$ may entail (spontaneous) loss of entanglement and loss of 
correlations between subsystems, but should not be able to create them.

\smallskip\noindent{\bf 10.\ Onsager reciprocity.}\label{reciprocity}
First, we introduce a particularly useful representation of general nonequilibrium states \cite{Beretta}. Given any state $\rho$ on $\Hil$, we define the effective Hilbert space $\Hil'$ as above, and choose a set  of operators $\{I',X'_1,X'_2,\dots\}$ spanning the linear space $\Ell_h(\Hil')$ of linear hermitian operators on $\Hil'$; the corresponding state $\rho'$ on $\Hil'$ has no zero eigenvalues, so that $S=-\Boltz B \ln\rho$ becomes $S'=-\Boltz \ln\rho'$ on $\Hil'$, which can be written as $ S'=f_0 I' +{\scriptstyle \sum_j} f_j X'_j$ because it belongs to $\Ell_h(\Hil')$. Thus,
\begin{equation}\label{anystate}
\rho' =\frac{ \exp (-\sum_j f_j  X'_j/\Boltz ) }{\Tr \exp (-\sum_j f_j  X'_j/\Boltz  )} \ ,
\end{equation}
where $f_0=\Boltz\ln\Tr \exp(-\sum_j f_j X'_j/\Boltz)$. Similarly, we can also write 
$S'_e=-\Boltz \ln\rho'_e =f_{0e} I' +{\scriptstyle \sum_j} f_{je} X'_j $, for the target maximum-entropy equilibrium state on $\Hil'$
\begin{equation} 
\rho'_e(\rho')=\frac{\exp(-\beta  H'+\sum_k \nu_k G'_k)} {\Tr\exp(-\beta  H'+\sum_k \nu_k G'_k)} \ ,
\end{equation}  
where $\beta$ and $\nu_k$ are such that $ e(\rho'_e)=e(\rho')$ and $ g_i(\rho'_e)=g_i(\rho')$, so that $\Tr(\rho'\ln\rho'_e)= \Tr(\rho'_e\ln\rho'_e)$ and $\Tr[({\rm d}\rho'/{\rm d}t)S'_e]=0$. As a result, the following relations can be easily proved,
\begin{eqnarray}
&s(\rho')- s(\rho'_e(\rho'))= f_0- f_{0e} + \sum_i (f_i- f_{ie})\, x_i (\rho') \ ,&\\
&\frac{\displaystyle \partial [s(\rho')- s(\rho'_e(\rho'))]}{ \displaystyle \partial x_i (\rho')}= f_i- f_{ie}\ ,&\label{affinity}\\
&\frac{\displaystyle {\rm d}s(\rho')}{\displaystyle {\rm d}t} = \sum_i f_i\frac{\displaystyle {\rm D} x_i(\rho')} {\displaystyle {\rm D}t}=\sum_i (f_i- f_{ie})\frac{\displaystyle {\rm D} x_i(\rho')} {\displaystyle {\rm D}t}\ ,&\\
&\cov{S'}{S'}=\sum_{ij} f_i f_j\cov{X'_i}{X'_j} \ , &
\end{eqnarray}
where ${\rm D}x_i(\rho')/{\rm D}t=\Tr (D_M X'_i)$ denotes the {\it dissipative rate of change} of the linear mean-value functional $x_i (\rho')=\Tr (\rho' X_i')$,
$\cov{S'}{S'}=\Tr [\rho'(-\Boltz \ln\rho')^2] -s(\rho')^2$, $\cov{X'_i}{X'_j} = \frac{1}{2}\Tr [(\rho\{ X'_i, X'_j\} ]- x_i(\rho') x_j(\rho')$.
When the system is in state $\rho'$, we interpret $ \cov{X'_i}{X'_j}$ as the codispersion (covariance) of simultaneous measurements of observables $X'_i$ and $X'_j$, $ \cov{X'_i}{X'_i}$ as the dispersion (or fluctuations) of observable $X'_i$ and $\cov{S'}{S'}$ the entropy fluctuations.

By Eq.\ (\ref{affinity}), we may also interpret $f_i- f_{ie}$ as the {\it generalized affinity} or {\it force} conjugated with the mean value of the linear observable $X_i$. In order to recover Onsager's theory, we may impose that (at least in the vicinity of state $\rho'_e$) the extended dynamics be such that the dissipative rates be linearly related to the generalized affinities through generalized-conductivity functionals, i.e.,
\begin{equation} 
\frac{\displaystyle {\rm D} x_i(\rho')} {\displaystyle {\rm D}t}={ \sum_j }L_{ij}(\rho',H',G'_k,X'_\ell,\dots) (f_j- f_{je}) \ ,
\end{equation} 
where the $ L_{ij}$'s may be nonlinear functionals of $\rho'$ (possibly to be approximated with their values at $\rho'_e$, in its vicinity), but should form a symmetric ($\bf{H}\rightarrow -\bf{H}$, if $H'$ depends on an external magnetic field) non-negative definite matrix, so that the rate of entropy production results in a quadratic form ${ \sum_{ij} }(f_i- f_{ie}) L_{ij} (f_j- f_{je})$.
Moreover,  the $ L_{ij}$'s should be linearly interrelated with the matrix of codispersions $ \cov{X'_i}{X'_j}$, in order to recover also Callen's fluctuation-dissipation theorem.


\begin{thebibliography}{16}
\bibitem{operatorB}
Given a $\rho$, $B=B^2= I- P_{\Ker(\rho)}= P_{\bot\Ker(\rho)}$, so that
$B=B^2$, $ B\rho=\rho B=\rho$,  and $B= B_A\Otimes B_B$ when $\rho= \rho_A\Otimes\rho_B$. Moreover, from $B=B^2$ follows that $B\dot B B=0$ and $\Tr(\rho \dot B \ln\rho)= \Tr(B\dot B B \rho \ln\rho)=0$, thus $\dot s(\rho)=\Tr (\dot\rho S)-\Boltz\Tr(\dot\rho)$ with $S=-\Boltz B\ln\rho$.
\bibitem{unitary}Under a unitary
(Hamiltonian) dynamical law the trajectories would be $u(t,\rho) =
U(t)\rho U^{-1}(t)$ with $U(t) = \exp(-itH/\hbar)$. The equilibrium states $\rho_e$, with $\rho_eH=H\rho_e$, would all be globally stable. Indeed, with
respect to the metric $d(\rho_1,\rho_2) = \Tr|\rho_1-\rho_2|$, it
is easy to show that every trajectory $u(t,\rho)$ would be
equidistant from any given equilibrium state $\rho_e$, i.e.,
$d(u(t,\rho),\rho_e) = d(u(0,\rho),\rho_e)$ for all $t$ and all
$\rho$. Because for each set of values of the invariant functianal these globally stable equilibrium states are in general more than one, the second-law
requirement of uniqueness would be violated.
\end{thebibliography}
\end{document}